\documentclass[aps,pre,twocolumn,groupedaddress,longbibliography]{revtex4-1}
\usepackage{graphicx} 
\usepackage{color}
\usepackage{times,mathptmx}
\usepackage{graphicx} 
\usepackage{amsmath}
\usepackage[normalem]{ulem}

\usepackage{cancel}

\begin{document}

\title{Buoyancy-driven dispersion in confined drying of liquid binary mixtures}
  
\author{Jean-Baptiste Salmon}
\affiliation{CNRS, Solvay, LOF, UMR 5258, Univ. Bordeaux, F-33600 Pessac, France.}

\author{Fr\'ed\'eric Doumenc}
\affiliation{Laboratoire FAST, Univ. Paris-Sud, CNRS, Universit\'e Paris-Saclay, F-91405, Orsay, France,}
\affiliation{Sorbonne Universit\'e, UFR 919, 4 place Jussieu, F-75252, Paris Cedex 05, France.}

\date{\today}

\begin{abstract}
We investigate the impact of buoyancy on the solute mass transport in an evaporating liquid mixture (non-volatile solute $+$ solvent) confined in a slit  perpendicular to the gravity.  Solvent evaporation at one end of the slit induces a solute concentration gradient which in turn drives free convection due to the difference between the densities of the solutes and the solvent. From the complete model coupling mass transport and hydrodynamics, 
we first  use a standard Taylor-like approach to  derive a one dimensional non-linear advection-dispersion equation  describing the solute concentration process for a dilute mixture. 
We then perform a complete analysis of the expected regimes using both scaling analysis  and asymptotic solutions of this equation. The validity of this approach is confirmed using a thorough comparison with the numerical resolution of both the complete model 
and the 1D advection-dispersion equation. Our results show that buoyancy-driven free convection always impacts solute mass transport at long time scales,
dispersing solutes in a steadily increasing length scale along the slit. 
Beyond this  confined drying configuration, our work also provides an easy way for evaluating the relevance of buoyancy on mass transport in  any other microfluidic configuration involving concentration gradients. 
\end{abstract}


\maketitle

\section{Introduction}

Drying of liquid mixtures  often leads to  concentration gradients, and therefore density gradients. 
When these gradients are orthogonal to gravity, they inevitably generate buoyancy-driven flows, that  then alters the drying process when coupled to the overall mass transport.
Solutal buoyancy-driven free convection is generally  relevant at relatively large scales, but many recent experiments reported such flows
in confined {\it microfluidic} geometries (10--100~$\mu$m): drying of confined~\cite{Daubersies:12,Lee:14,Pradhan2018,Loussert:16} and sessile droplets~\cite{Li2019,Kang2013,Edwards2018}, or any other microfluidic configuration generating concentration gradients~\cite{Gu2018,Selva:12}.
At small scales, buoyancy-driven flows are expected to play a minor role on mass transport, owing to the high viscous dissipation and fast solute diffusion~\cite{Squires:05}. Nevertheless, these flows always exist as soon as 
density gradients are perpendicular to gravity, and they can have an impact on less mobile species dispersed in the liquid mixture~\cite{Selva:12,Gu2018}.
\begin{figure}[ht]
\begin{center}
\includegraphics{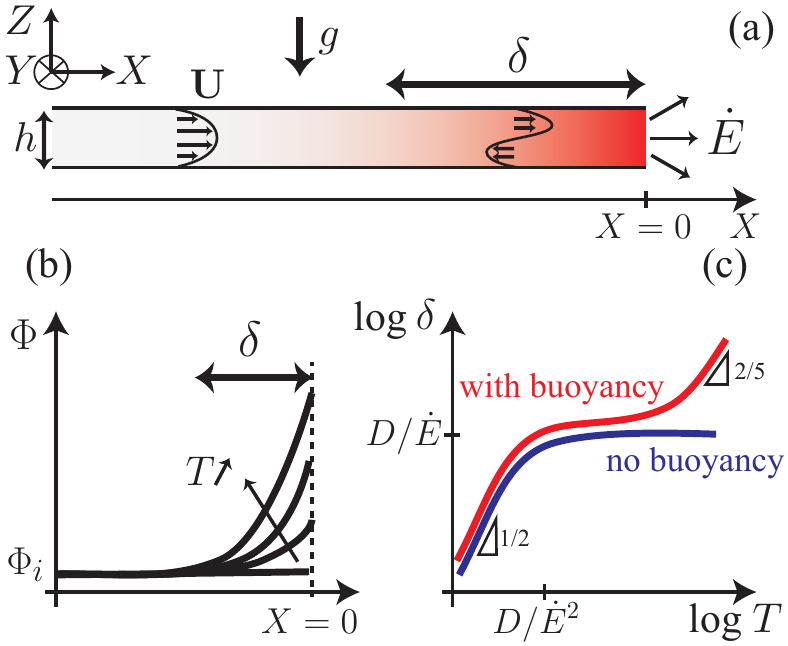}
\caption{(a) Schematic view of the confined drying experiment. The slit is  connected at $X \to -\infty$ to a tank containing the dilute mixture at concentration $\Phi_i$. (b) Solvent evaporation  at a rate $\dot{E}>0$ drives a flow concentrating continuously the non-volatile solute at the tip of the slit, see also the colored gradient in (a). 
The density gradient in turn generates a buoyancy-driven flow, superimposed on the evaporation-induced Poiseuille flow.
(c) Without buoyancy, $\delta$ initially grows as $\sqrt{D T}$ before reaching a steady value $\delta \sim D/\dot{E}$, whereas  the diffusive layer  invades the channel as $\delta \propto T^{2/5}$  when buoyancy-driven free convection dominates at long time scales. These two curves are slightly shifted for the sake of clarity. 
\label{fig:Evap1D}}
\end{center}
\end{figure}

The purpose of this work is to theoretically address such issues, and more precisely to quantitatively predict the range of parameters for which buoyancy-driven flows impact mass transport in a confined drying experiment. 
To do this, we consider the model experiment described in Fig.~\ref{fig:Evap1D}(a). 
A horizontal slit  of height $h$, initially filled with a liquid binary mixture,  is connected at one end to a tank containing the same mixture, and opened to the ambient atmosphere at the other end.
This geometry is not only prone to a simple modeling, but also commonly used to probe mass transport and uni-directional drying in complex fluids ranging from colloidal dispersions to surfactant mixtures, see e.g.~\cite{Dufresne:06,Inasawa:12,Boulogne:14,Lidon:14,Roger:16,Goehring2017}. In the following, we consider for simplicity a dilute binary mixture,  solvent $+$ non-volatile solute at concentration $\Phi$, for which both interdiffusion coefficient $D$ and kinematic viscosity $\nu$ are constant. 
Thereafter, we will also consider that the height of the slit is  small enough to neglect any inertial effect. 
Solvent evaporation, occurring at the outlet ($X=0$) at a rate $\dot{E}>0$  (m/s),  leads to a flow field $\mathbf{U}$ within the  slit. 
Owing to mass conservation, the horizontal component of the fluid velocity verifies 
\begin{eqnarray}
<U_X> = \frac{1}{h}\int_{0}^{h} U_X \text{d}Z = \dot{E}, \label{eq:convglobale}
\end{eqnarray}
 where $h$ is the height of the slit.
This flow continuously concentrates  the non-volatile solutes at the tip of  the slit, in a layer whose typical size $\delta$ depends on the competition between evaporation-induced advection and solute diffusion, see Fig.~\ref{fig:Evap1D}(b). 

Assuming first that diffusion homogenizes the concentration of solutes
over $h$, the solute concentration profile $\Phi(X,T)$ obeys the following 1D advection-diffusion equation:
\begin{eqnarray}
\frac{\partial \Phi}{\partial T} + \dot{E}\frac{\partial \Phi}{\partial X}= D \frac{\partial^2 \Phi}{\partial X^2}, \label{eq:trans1DdimFC}
\end{eqnarray} 
along with a solute no-flux boundary condition at $X=0$ as we consider non-volatile solutes. 
Fedorchenko and Chernov investigated theoretically this equation in the context of gas segregation induced by a moving solidification front~\cite{Fedorchenko2003} (see also Ref.~\cite{Sear2017} who used the same equation to describe stratification in drying films of colloidal dispersions).
According to Ref.~\cite{Fedorchenko2003}, solutes accumulate at the tip of the slit in a diffuse layer which first grows as $\delta \sim \sqrt{D T}$, and reaches the steady value $\delta \sim  D/\dot{E}$ after a transient time $\sim  D/\dot{E}^2$, see Fig.~\ref{fig:Evap1D}(c). In this asymptotic regime, the amount of solutes at the tip increases linearly with time, as well as the concentration gradient.

When the solute and solvent do not have the same density, such concentration gradients inevitably  generate buoyancy-driven free convection, see schematically Fig.~\ref{fig:Evap1D}(a).
Free convection is intrinsically a multi-dimensional (2D or 3D) problem.
In the present work, we will show that free convection due to buoyancy and its consequence on solute transport
in the slit sketched in Fig.~\ref{fig:Evap1D}  can be described by a  1D advection-dispersion equation. 
More precisely, we will use a standard Taylor-like approach  to show that the 
 transverse-averaged solute concentration profile defined by:
\begin{eqnarray}
\Phi_0(X,T) = \frac{1}{h}\int_{0}^h \Phi(X,Z,T) \text{d}Z,
\end{eqnarray}
is well approximated, for a wide range of parameters, by the solution of  
\begin{eqnarray}
\frac{\partial \Phi_0}{\partial T} + \dot{E}\frac{\partial \Phi_0}{\partial X}= \frac{\partial}{\partial X} \left(D_{\text{eff}} 
\frac{\partial \Phi_0}{\partial X}\right), \label{eq:trans1Ddim}
\end{eqnarray} 
where 
\begin{eqnarray}
 D_{\text{eff}} = D\left[1 + \frac{1}{\alpha}\left(\frac{g \beta_s h^4}{\nu D }\frac{\partial \Phi_0}{\partial X}\right)^2 \right], \label{eq:Deffdim}
\end{eqnarray} 
is a dispersion coefficient that takes into account both thermal diffusion and buoyancy on the solute mass transport. 
In the above equation, 
$g$ is the acceleration due to gravity,  $\beta_s$ the solutal expansion coefficient of the fluid mixture, and $\alpha = 362880$
($D$ is the interdiffusion coefficient and $\nu$ the kinematic viscosity).
 Chatwin and 
Erdogan were the first to derive this term when studying the effect of buoyancy on the dispersion  of solutes in a pressure-driven flow~\cite{Erdogan1967}. 
They also  reported that Taylor made the same calculation in 1953, but did not publish it, see also the  review of  Young
and Jones on shear dispersions~\cite{Young1991}.

The physics of Eqs.~(\ref{eq:trans1Ddim}-\ref{eq:Deffdim}) can be explained as follows.
In the framework of the lubrication approximation, the density gradient induces a flow $U_B$ whose scale comes from a balance between buoyancy ($g \beta_s h \frac{\partial\varphi_0}{\partial X}$) and viscous forces ($\nu U_B/h^2$), leading to~\cite{Young1991}: 
\begin{eqnarray}
U_B \sim \frac{g \beta_s h^3}{\nu}\frac{\partial \Phi_0}{\partial X}. \label{eq:UBscaling}
\end{eqnarray}
This flow, and more precisely its axial velocity distribution along $X$, see Fig.~\ref{fig:Evap1D}(a), increases solutes dispersion leading to  
the  term  in Eq.~(\ref{eq:Deffdim}). This term scales as $\sim (U_B h /D)^2$ thus comparing the solute diffusive transport ($D/h$) and its transport by buoyancy-driven convection ($U_B$). The prefactor $\alpha$, as well as the power $2$, comes from a standard Taylor-like perturbation approach, as first derived by Chatwin and 
Erdogan but in a circular tube~\cite{Erdogan1967}.
The non-linearity of this dispersive term, compared to the Taylor-Aris dispersion in a Poiseuille flow, comes from the coupling between the buoyancy-driven flow and the density gradient: strong gradients increase the magnitude of free convection, which in turn
increases solute dispersion,  see Eq.~(\ref{eq:UBscaling}).

Similar equations were also derived in various convection problems driven by temperature differences in a fluid layer, but also for describing the 
dynamics of well-mixed estuaries~\cite{Godfrey80,Smith1976}, for quantifying the impact of buoyancy on the measurement of diffusivities  in liquid metals~\cite{MACLEAN:01}, or even
to study gravity currents of miscible fluids in porous media~\cite{Szulczewski2013}. 
Despite an in-depth literature review on this classical Taylor-like approach, we are not aware of any work discussing such an equation in the context of confined drying, and more generally of microfluidic experiments generating solute gradients.

In a second step, we thus  report a complete investigation of the solute concentration process described by the advection-dispersion equation Eq.~(\ref{eq:trans1Ddim}). 
We  first predict all the expected regimes of solute concentration using a detailed scaling analysis, and we then provide asymptotic analytical solutions of the concentration profiles
  when mass transport is dominated by diffusion or else by buoyancy-driven dispersion. 

In particular, we show that solutes dispersion caused by buoyancy   leads at long time scales   to a diffusive  layer thickness $\delta$ which continuously invades the channel following $\delta \propto T^{2/5}$, see Fig.~\ref{fig:Evap1D}(c).
  This theoretical approach also helps us  to provide  a simple diagram highlighting the range of parameters  corresponding to mass transport dominated by diffusion (i.e. negligible dispersion). 	
Finally, the validity of the 1D advection-dispersion model is thoroughly investigated by means of scaling analysis, and the results confirmed by comparisons with direct numerical simulations of the 2D model.

The present paper is organized as follows. In Section~\ref{sec:model}, we first present the equations modeling the experiments shown in Fig.~\ref{fig:Evap1D}, as well as 
the underlying assumptions.  We then derive from this model the advection-dispersion equation Eq.~(\ref{eq:trans1Ddim}).
Section~\ref{sec:sol} then reports  a complete discussion of the different regimes expected, as well as the corresponding analytical asymptotic solutions. The validity of the 1D approach is then investigated. In Section~\ref{sec:conclusion}, we finally conclude our work by discussing its possible implications, particularly for microfluidic experiments generating concentration gradients.

\section{From the 2D model to a 1D advection-dispersion equation \label{sec:model}}

\subsection{2D model}
As stated in the introduction, we first consider a binary mixture, solvent $+$ non-volatile solute with concentration $\Phi$. We also assume that its density evolves as:
\begin{eqnarray}
\rho = \rho_i[1+ \beta_s (\Phi-\Phi_i)]\,,
\end{eqnarray}
where $\rho_i$ is the density at the solute concentration $\Phi_i$, and $\beta_s$ the solutal expansion coefficient of the fluid mixture at the reference concentration $\Phi_i$.
The configuration under study is depicted in Fig.~\ref{fig:Evap1D}, and we consider that both the height of the slit  and the evaporation-induced flow are small enough to neglect inertia. The  slit is initially filled homogeneously by the solution at concentration $\Phi_i$.
Solvent evaporation induces the concentration of the non-volatile solutes at the tip of the  slit. Thereafter, we limit our study to dilute solutions, for which the evaporation rate $\dot{E}$ and the various
transport coefficients (viscosity $\nu$, mutual diffusion coefficient $D$) remain constant. The following model, coupling Stokes, continuity, and solute conservation equations, is expected to describe the overall concentration process:
\begin{eqnarray}
&&\rho_i\nu\Delta \mathbf{U} - \nabla P + (\rho(\Phi)-\rho_i) \mathbf{g}= 0, \label{eq:stokdim}\\
&&\nabla . \mathbf{U} = 0, \label{eq:contdim}\\
&&\partial_T \Phi + \mathbf{U}.\nabla \Phi = D \Delta \Phi, \label{eq:convdim}
\end{eqnarray}
where $\mathbf{U}$ is the velocity field of the mixture, and $P$ the pressure deviation from the initial hydrostatic pressure field.  

We also assume the following standard boundary conditions at the solid walls:
\begin{eqnarray}
&&\mathbf{U}  = 0~~\text{(no slip),} \label{eq:slip}\\
&&\nabla \Phi .\mathbf{n} = 0~~\text{(no flux),} \label{eq:adiab}
\end{eqnarray}
and the following standard  ones at the evaporating free surface:
\begin{eqnarray}
&&U_X(X=0,Z,T) = \dot{E}, \label{eq:afreesurface} \\
&&\left(\frac{\partial U_Z}{\partial X}\right)(X=0,Z,T)   = 0, \label{eq:bfreesurface} \\
&&\left(U_X \Phi - D \frac{\partial \Phi}{\partial X}\right)(X=0,Z,T) = 0. \label{eq:cfreesurface}
\end{eqnarray}
Eq.~(\ref{eq:cfreesurface}) ensures the non-volatility of the solute.
At $X \to -\infty$, we impose:
\begin{equation}
	\Phi(X \to -\infty ,Z,T) = \Phi_i. \label{eq:CLinf}
\end{equation}

\subsection{Numerical resolution on a given experimental case \label{experimentalcase}}
To illustrate the impact of  buoyancy,  we first consider the following  realistic case: water evaporation from an aqueous dispersion of silica nanoparticles at ambient conditions in  a slit of height $h=150~\mu$m.  
Similar  conditions were recently explored experimentally either to probe mass transport in such charged dispersions~\cite{Goehring2017,Sarkar:09,Lidon:14}, or to investigate the dynamics of fractures, delamination and shear bands in the concentrated regime, see e.g.~\cite{Dufresne:06,Inasawa:12,Boulogne:14}.
We will assume an initial concentration $\Phi_i = 0.001$ and radius $a= 5$~nm for the silica nanoparticles, leading to a 
diffusivity $D \simeq 4.37 \times 10^{-11}$~m$^2$/s according to the Stokes-Einstein relation for a temperature of 25$^\circ$C (we do not consider here enhanced values due to colloidal interactions occurring for such systems at high concentrations~\cite{Goehring2017,Loussert:16}). For such a very dilute silica dispersion, the solutal expansion coefficient at the reference concentration $\Phi_i$ is well approximated by $\beta_s \simeq \rho_s/\rho_w -1 \simeq 1.2$ where $\rho_s \simeq 2200$~kg/m$^3$ is the density of silica, and  $\rho_w \simeq 1000$~kg/m$^3$ that of water. 
We assume an  evaporation rate $\dot{E} = 0.1~\mu$m/s. The latter remains constant as the volume of the colloids is much larger than the water molecular volume~\cite{Russel}. 
We will finally assume that the kinematic viscosity $\nu = 10^{-6}$~m$^2$/s remains also constant during concentration.

The numerical resolution of Eqs.~(\ref{eq:stokdim}--\ref{eq:convdim}) with boundary conditions Eqs.~(\ref{eq:slip}--\ref{eq:cfreesurface}) has been performed with the commercial software Comsol Multiphysics (finite elements, Galerkin method).
The boundary condition Eq.~(\ref{eq:CLinf}) at $X \to -\infty$ has been moved to $X=-L$, where $L = 10$~mm is a finite distance large enough to not affect the results significantly  
(in addition, the pressure $P$ has been arbitrarily set to $P=0$ at $X=-L$).
Time discretization is based on implicit backward differentiation formulas.
Spatial discretization was achieved by a structured mesh of quadratic Lagrangian elements. 
The mesh convergence has been thoroughly tested by successive refinements.

\begin{figure}[ht]
\begin{center}
\includegraphics{{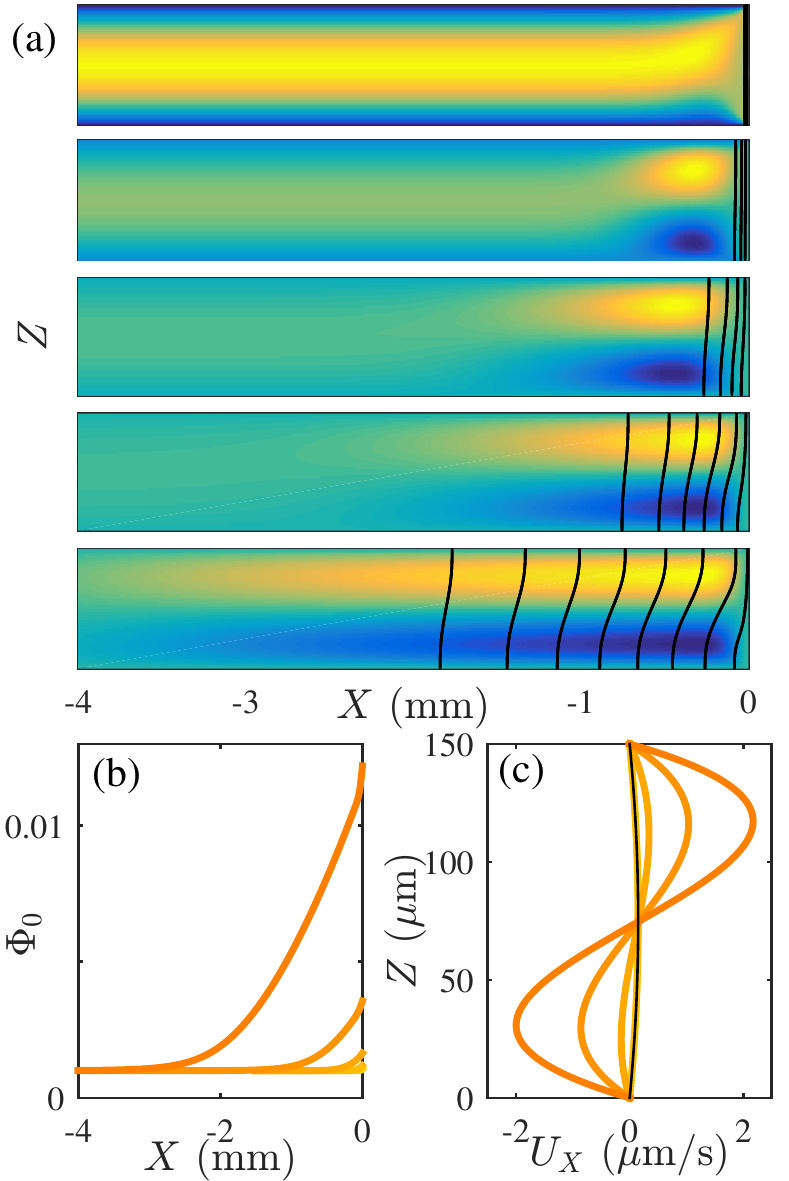}}
\caption{(a) Snapshots of $U_X$ (colors) and $\Phi(X,T)$ (contours) at $T=10$, $10^2$, $10^3$, $10^4$, and $10^5$~s, from top to bottom. The range of the colormap is scaled by the maximal velocity in each plot (b) Height-averaged concentration profiles $\Phi_0(X,T)$ and (c)
velocity profiles $U_X(Z,T)$ at $X=-300~\mu$m, at the  same times $T$ as in (a). The thin dark line is the evaporation-driven Poiseuille profile $U_P=6\dot{E}Z(h-Z)/h^2$.
\label{fig:FigureFD1}}
\end{center}
\end{figure}

Figure~\ref{fig:FigureFD1} shows the results of the numerical simulation.
More precisely, Fig.~\ref{fig:FigureFD1}(a) shows several snapshots at $T=10$, $10^2$, $10^3$, $10^4$, and $10^5$~s of both the horizontal component $U_X$ of the velocity field (colormap) and the concentration profile $\Phi(X,T)$ (contour). Figures~\ref{fig:FigureFD1}(b) and (c) display  the height-averaged concentration profiles $\Phi_0$ and  $U_X$ at $X = -300~\mu$m at the same times.
For the conditions investigated, the velocity profile $U_X(Z,T)$ at $X= -300~\mu$m and $T=10~$s mainly corresponds  to the evaporation-driven Poiseuille flow $U_P=6\dot{E}Z(h-Z)/h^2$. However the profiles are more and more distorted at longer times by the free convection induced by the solute concentration gradient. The maximal velocity at $X=-300~\mu$m due to buoyancy only, i.e. $U_B = U-U_P$, increases from $\simeq 0.24~\mu$m/s at $T=10^3$~s to $\simeq  2~\mu$m/s at $T=10^5~$s. As clearly evidenced by the contour plot in  Fig.~\ref{fig:FigureFD1}(a), buoyancy also distorts the isoconcentration lines, and therefore clearly impacts the solute mass transfer.
Note that this 2D model predicts that the concentration  at the interface increases continuously, and the assumption of constant kinematic viscosity and mutual diffusion coefficient may not hold anymore above $T>10^5$~s for which the concentration  at the interface reaches $\simeq 0.012$.

\subsection{Advection-dispersion equation in the framework of the lubrication approximation \label{sec:dimensionless}}

As discussed in the introduction, our aim is to predict the range of parameters for which buoyancy impacts solute's transport in such confined drying configuration using 
a 1D advection-dispersion equation derived from the above model.
As shown  by Fedorchenko and Chernov~\cite{Fedorchenko2003},  $D/\dot{E}$ is a length scale  that naturally emerges along $X$ from the conservation equation Eq.~(\ref{eq:convdim}), and we therefore define the 
following dimensionsless  variables:
\begin{eqnarray}
&&z  =  Z/h,~~x =  (\dot{E}/D) X,~~t = (\dot{E}^2/D)T, \notag\\
&& u_X =  U_X/\dot{E},~~u_z = U_Z/(\dot{E} \text{Pe}),\notag\\
&&\varphi = (\Phi-\Phi_i)/\Phi_i,~~p = h^2/(\rho_i \nu D) P, \label{eq:unitless}
\end{eqnarray} 
where $\text{Pe}$ is the P\'eclet number given by:
\begin{eqnarray}
\text{Pe}  = \frac{\dot{E}h}{D}.
\end{eqnarray} 
Note that we have chosen two different scales to obtain the dimensionless coordinates $x$ and $z$. As shown later, this particular choice makes it easy to highlight
the different regimes of solute concentration, while keeping a compact writing of the equations. However, it should be remembered that the scale $D/\dot{E}$ is implicitly contained in the dimensionless abscissa $x$ but not in $z$.

Using this set of dimensionless variables,  one can demonstrate that the dimensionless counterpart of  Eqs.~(\ref{eq:stokdim}-\ref{eq:convdim})   
 depend only on two parameters, Pe and the solutal Rayleigh number defined as:
 \begin{eqnarray}
\text{Ra}  = \frac{\beta_s \Phi_i g h^3}{\nu D}, \label{eq:defRa}
\end{eqnarray} 
see Appendix~\ref{app:derivation}.
Volume integration of the continuity relation Eq.~(\ref{eq:contdim}) yields the dimensionless counterpart of the solution global mass balance Eq.~(\ref{eq:convglobale}):
\begin{eqnarray}
<u_x> = \int_{0}^1 u_x \text{d}z = 1. \label{eq:convglobaleadim}
\end{eqnarray}
Similarly, integration of Eq.~(\ref{eq:convdim}) yields the dimensionless solute global mass balance:   
\begin{eqnarray}
\int_{x \to -\infty}^0  \int_{z=0}^1 \varphi  \text{d}z  \text{d}x =  t.  \label{eq:convglobalsoluteeadim}
\end{eqnarray}

 We now assume that the typical scale $\delta$ of both the concentration gradient and the buoyancy-driven velocity field, see Fig.~\ref{fig:Evap1D}, is much larger than the channel height, i.e. $\delta \gg \text{Pe}$ with our dimensionless variables, Eqs.~(\ref{eq:unitless}). 
We therefore assume quasi-parallel flows and we use the standard lubrication approximation~\cite{Oron:97} to derive a 1D solute conservation equation from the above model. 
 More precisely, we use a standard Taylor-like perturbation method,  as reviewed for instance in Young and Jones' work on shear dispersion~\cite{Young1991}, and expand the
concentration field as:
\begin{eqnarray}
\varphi(x,z,t) = \varphi_0(x,t) + \text{Pe}^2 \varphi_1(x,z,t), \label{eq:perturbation}
\end{eqnarray}
where $\varphi_0(x,t)$ is the transverse-averaged concentration profile, i.e. $\varphi_0(x,t) = <\varphi(x,z,t)>$ with the same averaging as in Eq.~(\ref{eq:convglobaleadim}), and $\text{Pe}^2 \varphi_1 \ll \varphi_0$.
Appendix~\ref{app:derivation} presents the detailed derivation leading ultimately to the following transport equation for the mean concentration field $\varphi_0$:
\begin{eqnarray}
&&\frac{\partial  \varphi_0}{\partial t} + \frac{\partial  \varphi_0}{\partial x}  =  \frac{\partial}{\partial x}\left( D_{\text{eff}} \frac{\partial \varphi_0}{\partial x} \right), 
\label{eq1Daverage}
\end{eqnarray}
assuming $t \gg \text{Pe}^2$, $\delta \gg \text{Pe}$, and $\text{Pe}^2 \varphi_1 \ll \varphi_0$~\cite{Young1991}.
The range of validity of these conditions will be discussed later on  in Sec.~\ref{sec:validity}.

The dispersion coefficient $D_{\text{eff}}$ is given by:
\begin{multline}
D_{\text{eff}} = 1 + \frac{(\text{Pe}\text{Ra})^2}{\alpha}\left(\frac{\partial  \varphi_0}{\partial  x}\right)^2 +  \beta \text{Pe}^2 \\ \simeq  1 + \frac{(\text{Pe}\text{Ra})^2}{\alpha}\left(\frac{\partial  \varphi_0}{\partial  x}\right)^2
\label{eq:Deffsansapp}
\end{multline}
where $\beta = 1/210$ and $\alpha = 362880$.
The first term corresponds to the  dispersion of the solutes due to the axial velocity distribution caused by the density gradient, as discussed in Introduction, whereas the second is the traditional Taylor-Aris term induced by the Poiseuille flow. 
There are no coupling terms in the geometry under study, between buoyancy and Taylor-Aris dispersion, owing to the symmetry along the plane $z=1/2$, as already noted in~\cite{Young1991}.
Eqs.~(\ref{eq1Daverage}-\ref{eq:Deffsansapp}) finally lead to  Eqs.~(\ref{eq:trans1Ddim}-\ref{eq:Deffdim}) with real units as   we will assume that the Taylor-Aris term is negligible in our configuration, i.e. $\beta\text{Pe}^2 \ll 1$, see Sec.~\ref{sec:validity} regarding
 the range of validity of our model. 
Note  importantly that despite the fact that the dispersion term due to buoyancy in Eq.~(\ref{eq:Deffsansapp}) depends explicitly on the P\'eclet number $\text{Pe}$ in our dimensionless model, the latter actually does not depend on the evaporation rate $\dot{E}$  due to the scale $D/\dot{E}$ used to define  $x$, see Eqs.~(\ref{eq:unitless}).

It should also be noted  that many groups recently investigated the drying of droplets confined between two circular parallel plates, see e.g.~\cite{Selva:12,Daubersies:12,Lee:14,Pradhan2018,Loussert:16}. Most of these works clearly reported buoyancy-driven flows generated by the radial density gradients induced by solvent evaporation. 
Interestingly, the same above calculations applied to this cylindrical geometry and 
for a binary mixture, lead also to Eqs.~(\ref{eq1Daverage}-\ref{eq:Deffsansapp}) with $\alpha = 362880$ but without the Taylor-Aris and advection terms and with cylindrical
coordinates. 
Our theoretical derivation does therefore not only apply to the case of  a slit, but also to this specific 2D configuration.

\section{Dynamics of the solute concentration\label{sec:sol}}

In the following, we now turn to a thorough analysis of the solute concentration process in the confined drying experiment described in Fig.~\ref{fig:Evap1D}. More precisely, we will  use the 1D advection-dispersion equation derived above, Eq.~(\ref{eq1Daverage}), to unveil both the expected regimes of solute concentration and the relevant parameters impacting the solute mass transport.
Eq.~(\ref{eq1Daverage}) is supplemented with the following initial and boundary conditions:
\begin{eqnarray}
&&\varphi_0(x,t=0) = 0, \label{eq:phicl2}\\
&&\lim_{x \to - \infty} \varphi_0(x,t) = 0, \label{eq:phicl1}\\
&&\left(1+\varphi_0-D_{\text{eff}}\frac{\partial \varphi_0}{\partial x}\right)(x=0,t), \label{limeq1Daverage}
\end{eqnarray}
where Eq.~(\ref{limeq1Daverage}) ensures  the condition of zero solute flux through the free surface (non volatility of the solute).

In the framework of this 1D model, the global solute mass balance Eq.~(\ref{eq:convglobalsoluteeadim}) reads:
\begin{eqnarray}
\int_{x \to -\infty}^0  \varphi_0(x,t) \text{d}x = t.  \label{eq:convglobalsoluteeadim2}
\end{eqnarray}	
Finally, an important feature of the concentration field is its spatial extent $\delta$ as illustrated in Fig.~\ref{fig:Evap1D}, and we define the latter according to:
 \begin{eqnarray}
\delta(t) = -\frac{\int_{-\infty}^0 x\varphi_0(x,t)\,\text{d}x}{\int_{-\infty}^0 \varphi_0(x,t)\,\text{d}x} = -\frac{1}{t}\int_{-\infty}^0 x\varphi_0(x,t)\,\text{d}x. 
\label{eq:defdelta}
\end{eqnarray}

\subsection{Scaling analysis\label{sec:scalinganalysis}}
We first use a standard method~\cite{bejan95} to derive the scaling laws of the model defined by the governing equation Eq.~(\ref{eq1Daverage}), along with the initial and boundary conditions  Eq.~(\ref{eq:phicl2}--\ref{limeq1Daverage}). 
The global mass balance Eq.~(\ref{eq:convglobalsoluteeadim2}), which implicitly contains the governing equation and its boundary and initial conditions, provides a first scaling law:
\begin{equation}
\varphi_0 \delta \sim t. \label{eq:ScaleGlob}
\end{equation}

A second scaling law is provided by Eq.~(\ref{eq1Daverage}), which can be written as  a relation between 4 positive terms:
\begin{equation} \label{eq:phigov}
\frac{\partial \varphi_0 }{\partial t} + \frac{\partial \varphi_0 }{\partial x} =  \frac{\partial^2 \varphi_0 }{\partial x^2} 
+ \frac{(\mathrm{Pe Ra})^2}{ \alpha}  \frac{\partial}{\partial x}  \left(\frac{\partial \varphi_0}{\partial x} \right )^3.
\end{equation}
Owing to the boundary condition Eq.~(\ref{eq:phicl1}), the order of magnitude of these four terms is:
\begin{equation}
\frac{\varphi_0}{t} ~ ~ ~ ; ~ ~ ~ \frac{\varphi_0}{\delta}  ~ ~ ~ ; ~ ~ ~ \frac{\varphi_0}{\delta^2} ~ ~ ~ ; ~ ~ ~  \frac{(\text{Pe}\text{Ra})^2}{\alpha}\frac{\varphi_0^3}{\delta^4}.  \label{eq:Scal}
\end{equation}
As time goes by, different regimes are encountered depending on what couple of terms dominates in Eq.~(\ref{eq:phigov}).
As all the terms of this equation are positive, 
the balance of the different orders of magnitude reads:
\begin{equation} \label{eq:regimes}
\max { \left ( \frac{\varphi_0}{t} ~ , ~ \frac{\varphi_0}{\delta} \right) } \sim  \max { \left ( \frac{\varphi_0}{\delta^2} ~ , ~ \frac{(\text{Pe}\text{Ra})^2}{\alpha}\frac{\varphi_0^3}{\delta^4} \right) },
\end{equation}
and one therefore expects four different regimes, systematically reviewed in the following.

\paragraph{Diffusive regime D1 ---}
We define this regime as the one corresponding to $ \frac{\varphi_0}{t} \sim \frac{\varphi_0}{\delta^2}$ in Eq.~(\ref{eq:regimes}). 
Combining this relation with Eq.~(\ref{eq:ScaleGlob}) yields:
\begin{eqnarray} \label{eq:D1}
&&\delta \sim \sqrt{t}~~\text{and}~~\varphi_0 \sim \sqrt{t},\\
&& t \ll 1~~\text{and}~~\frac{\text{Pe}\text{Ra}}{\sqrt{\alpha}}\ll 1,  \label{eq:D1bis}
\end{eqnarray}
\paragraph{Diffusive regime D2 ---}
This regime corresponds to $\frac{\varphi_0}{\delta} \sim \frac{\varphi_0}{\delta^2}$ and thus to
\begin{eqnarray} \label{eq:D2}
&&\delta \sim 1~~\text{and}~~\varphi_0 \sim t,\\
&& t \gg 1~~\text{and}~~t \ll \left(\frac{\text{Pe}\text{Ra}}{\sqrt{\alpha}}\right)^{-1}, \label{eq:D2bis}
\end{eqnarray}
and D2 can thus only be observed  when $\text{Pe}\text{Ra} \ll \sqrt{\alpha}$
\paragraph{Dispersive regime C1 ---}
This regime corresponds to $\frac{\varphi_0}{t} \sim \frac{(\text{Pe}\text{Ra})^2}{\alpha}\frac{\varphi_0^3}{\delta^4}$, and after calculation to
\begin{eqnarray} \label{eq:C1}
&&\delta \sim \left(\frac{\text{Pe}\text{Ra}}{\sqrt{\alpha}}\right)^{1/3} \sqrt{t}~~\text{and}~~\varphi_0 \sim \left(\frac{\text{Pe}\text{Ra}}{\sqrt{\alpha}}\right)^{-1/3} \sqrt{t},\\
&& t \ll \left(\frac{\text{Pe}\text{Ra}}{\sqrt{\alpha}}\right)^{2/3}~~\text{and}~~\left(\frac{\text{Pe}\text{Ra}}{\sqrt{\alpha}}\right) \gg 1.
\end{eqnarray}
\paragraph{Dispersive regime C2 ---}
This last regime corresponds to $\frac{\varphi_0}{\delta}  \sim \frac{(\text{Pe}\text{Ra})^2}{\alpha}\frac{\varphi_0^3}{\delta^4}$, leading after calculation to 
\begin{eqnarray} \label{eq:C2}
&&\delta \sim \left(\frac{\text{Pe}\text{Ra}}{\sqrt{\alpha}}\right)^{2/5} t^{2/5}~~\text{and}~~\varphi_0 \sim \left(\frac{\text{Pe}\text{Ra}}{\sqrt{\alpha}}\right)^{-2/5} t^{3/5},\\
&& t \gg \left(\frac{\text{Pe}\text{Ra}}{\sqrt{\alpha}}\right)^{2/3}~~\text{and}~~t \gg \left(\frac{\text{Pe}\text{Ra}}{\sqrt{\alpha}}\right)^{-1}.\label{eq:C22}  \label{eq:C2bis}
\end{eqnarray}

The above scaling analysis reveals the importance of the dimensionless parameter $\text{Pe}\text{Ra}$. 
When $\text{Pe}\text{Ra} \ll \sqrt{\alpha}$, the sequence of regimes is D1 $\to$ D2 $\to$ C2 with two transition times $t_{\text{D1} \to \text{D2}} \sim 1$ and
\begin{align}
t_{ \text{D2} \to \text{C2}} \sim \left ( \frac {\text{Pe}\text{Ra}} {\sqrt{\alpha}} \right )^{-1}.  \label{eq:tD2C2}
\end{align}
For this range of $\text{Pe}\text{Ra}$,  we found at small time scales the classical scenario of Fedorchenko and Chernov~\cite{Fedorchenko2003}  expected without buoyancy and displayed in Fig.~\ref{fig:Evap1D}(c): 
the square-root growth $\delta \sim \sqrt{t}$ ($\delta \sim \sqrt{DT}$ with real units) of a diffusive layer reaching after a transient 
$t \sim 1$ ($T \sim D/\dot{E}^2)$, a constant value $\delta \sim 1$ ($\delta \sim D/\dot{E}$) owing to the competition between solute diffusion and evaporation-driven advection. On longer time scales, however, the solute concentration gradient steadily increases, and buoyancy can no longer be ignored (transition D2 $\to$ C2). 

When $\text{Pe}\text{Ra} \gg \sqrt{\alpha}$,  dispersion caused by buoyancy always dominates diffusion in the solute mass transport,  and the sequence of observed regimes reduces to C1 $\to$ C2 with a transition at $t_{\text{C1} \to \text{C2}} \sim  (\text{Pe}\text{Ra}/\sqrt{\alpha})^{2/3}$. Importantly, this scaling analysis also demonstrates that solutes are always dispersed at long time scales in a steadily increasing diffusive layer following $\delta \propto t^{2/5}$. 

\subsection{Numerical simulation of the 1D advection-dispersion model and asymptotic solutions}
\begin{figure}[ht]
\begin{center}
\includegraphics{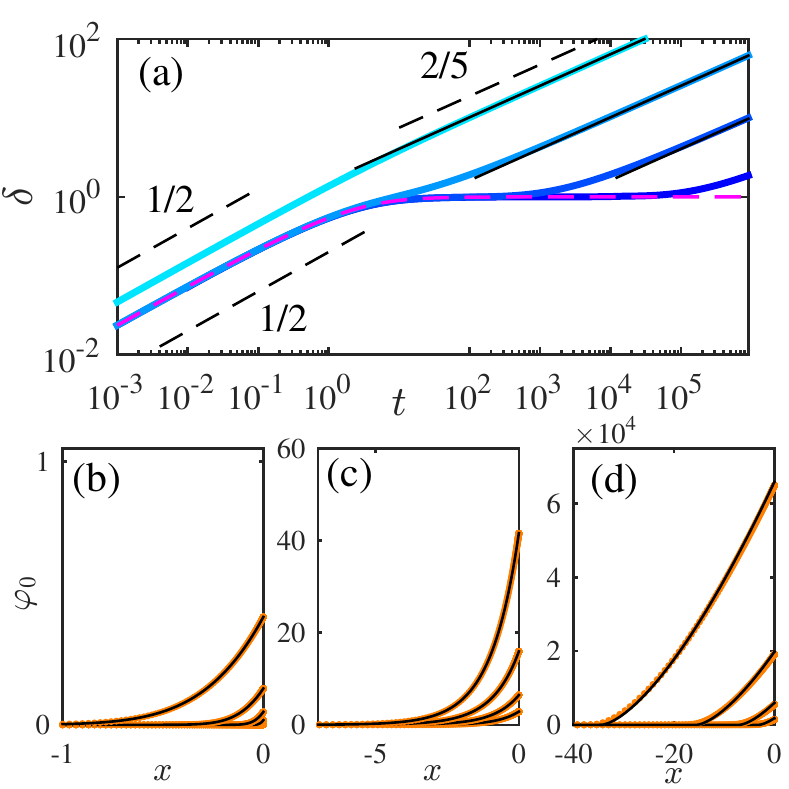}
\caption{(a) Numerical solution $\delta$ vs. $t$ of the 1D advection-dispersion model Eqs~(\ref{eq1Daverage}--\ref{limeq1Daverage})  for different $\text{Pe}\text{Ra}$: $10^{-2}$, $10^0$, $10^{2}$, and   $10^4$  (from dark to light blue).
The magenta dashed line is the theoretical prediction Eq.~(\ref{eq:analdelta}) in the  diffusive  regimes D1 and D2. The black lines are the approximations Eq.~(\ref{eq:deltadiffbuo}) for the dispersive  regime C2. (b) to (d): $\varphi(x,t)$ vs. $x$ computed from  Eqs~(\ref{eq1Daverage}--\ref{limeq1Daverage}) at $\text{Pe}\text{Ra}=1$.
(b) regime D1: $t<0.5$; (c) D2: $1<t<100$; and (d) C2:  $t>1000$. In (b) and (c), the analytical solutions given by Eq.~(\ref{eq:Phi0analytique}) (black lines) are superimposed with the numerical resolution. In (d), black lines correspond to  the approximate solution Eq.~(\ref{eq:appIPhi}).   
\label{fig:Evap1Deff_pdepe}}
\end{center}
\end{figure}
To illustrate the above scaling analysis, 
Fig.~\ref{fig:Evap1Deff_pdepe}(a) displays the temporal evolution of the thickness of the diffusive layer $\delta$, calculated from the numerical resolution of Eqs.~(\ref{eq1Daverage}--\ref{limeq1Daverage}) for a wide range of $\text{Pe}\text{Ra}$ ranging logarithmically from $10^{-2}$ to $10^4$. 

Except for  $\text{Pe}\text{Ra}\geq \sqrt{\alpha}$, these curves display three regimes:  an initial growth of the diffusive layer following $\delta \sim \sqrt{t}$ (regime D1),  a constant plateau $\delta \simeq 1$ reached at $t\simeq 1$ (regime D2), followed again by the growth 
of $\delta$ at longer time scales according to $\delta \propto t^{2/5} $(regime C2). 
The departure from the regime D2 to the regime C2 occurs at a critical time which decreases for increasing $\text{Pe}\text{Ra}$.
For $\text{Pe}\text{Ra} \geq \sqrt{\alpha}$, the  regime D2 of constant diffusive layer does not exist, and the initial regime C1 for which we observe again $\delta \propto \sqrt{t}$, 
does not collapse with the other curves in the regime D1.
Figures~\ref{fig:Evap1Deff_pdepe}(b-d) illustrate the specific case $\text{Pe}\text{Ra} = 1$, for which the sequence of the three different regimes D1 $\to$ D2 $\to$ C2 is easily revealed.  

All these numerical results are obviously in line with the scaling analysis reported in Sec.~\ref{sec:scalinganalysis}, but one should go beyond the scaling laws
Eqs.~(\ref{eq:D1}--\ref{eq:C22}) in order to predict quantitatively the range of parameters for which  the solute mass transport is dominated by diffusion only.

When diffusion dominates  mass transport (regimes D1 and D2), Eq.~(\ref{eq1Daverage}) reduces to the following linear advection-diffusion equation:
\begin{eqnarray}
&&\frac{\partial \varphi_0}{\partial t} + \frac{\partial \varphi_0}{\partial x}  = \frac{\partial^2 \varphi_0}{\partial x ^2}. \label{eq:transport1D}
\end{eqnarray}
The analytical solution of this advection-diffusion equation  has been derived by Fedorchenko and Chernov~\cite{Fedorchenko2003} see Eq.~(\ref{eq:Phi0analytique}) in Appendix~\ref{appsolutionanalytik}.
This analytical solution superimposes perfectly on the profiles computed numerically from Eq.~(\ref{eq1Daverage}) at small time scales, see Fig.~\ref{fig:Evap1Deff_pdepe}(b) and (c). We also computed $\delta$ vs. $t$ from this analytical solution:
\begin{multline} 
\delta(t) = \frac{-t^2 + \left[t(4+t)-4\right]\text{erf}\left(\frac{\sqrt{t}}{2}\right)}{4 t }  \\
+ \frac{(2+t)\sqrt{t}\exp\left(-\frac{t}{4}\right)}{2\sqrt{\pi}t }, \label{eq:analdelta}
\end{multline}
see the magenta dashed line  in Fig.~\ref{fig:Evap1Deff_pdepe}(a). 

In the  dispersive regime C2, we used the integral method~\cite{Ozisik} to approximate the concentration profiles.
The detailed calculation and the resulting approximations are  given in Appendix~\ref{appMethoInt}, see in particular Eq.~(\ref{eq:appIPhi})
from which we computed $\delta(t)$:
\begin{eqnarray}
\delta(t) = \frac{5^{2/5}3^{3/5}}{7}\left( \frac{\text{Pe} \text{Ra}}{\sqrt{\alpha}}  \, t \right)^{2/5}. \label{eq:deltadiffbuo}
\end{eqnarray} 
 Both  $\delta$ vs. $t$ and concentration profiles perfectly 
superimpose on the solutions computed numerically, see Fig.~\ref{fig:Evap1Deff_pdepe}(a) and Fig.~\ref{fig:Evap1Deff_pdepe}(d) respectively.

We did not find an approximate solution of the concentration profiles in the  dispersive regime C1. Nevertheless, we computed numerically the prefactor in Eq.~(\ref{eq:C1}) leading to:
\begin{equation}
\delta(t)  \simeq 0.55 \left( \frac{\text{Pe} \text{Ra} }{\sqrt{\alpha}}\right)^{1/3} \sqrt{t}. \label{eq:deltaC2}
\end{equation}

\subsection{Transition times} \label{TransTime}

We define a criterion for evaluating quantitatively the critical time corresponding to the transition from a diffusive regime, for which 
buoyancy-driven dispersion  exists but has a negligible impact on the solute transport, 
to  a  dispersive \ regime, for which  buoyancy significantly disperses solutes along the slit.
We consider that this transition occurs when the ratio of the dispersion flux to the height-averaged diffusive flux reaches a given value $p$. This condition reads:
\begin{equation}
	\frac {< u_x ·\varphi > - \varphi_0 } { \frac{\partial \varphi_0}{\partial x}} = p, 	\label{eq:crit}
\end{equation}
at a given abscissa $x$. 
We arbitrarily set $p=0.2$ for all the calculations in the present work. In the framework of the 1D advection-dispersion model, Eq.~(\ref{eq:crit}) turns to
\begin{equation}
	D_{\text{eff}} = 1 + p. 	\label{eq:crit1D}
\end{equation}
Using the expression of the dispersion coefficient Eq.~(\ref{eq:Deffsansapp}) 
and considering that the concentration gradient is maximum at $x=0$, Eq.~(\ref{eq:crit1D}) reads  
\begin{eqnarray} \label{eq:Deffapp}
\text{Pe} \text{Ra} \left(\frac{\partial \varphi_0}{\partial x}\right)_{x=0,~t=t_{ \text{D2} \to \text{C2}}} =  \sqrt{p \alpha}.
\end{eqnarray}
The critical time $t_{\text{D2} \to \text{C2}}$ is computed by numerically solving Eq.~(\ref{eq:Deffapp}), 
where the concentration gradient  at  $x=0$ is estimated from the analytical expression Eq.~(\ref{eq:Phix0TEXT}), valid in the diffusive regimes D1 and D2.
\begin{figure}[ht]
\begin{center}
\includegraphics{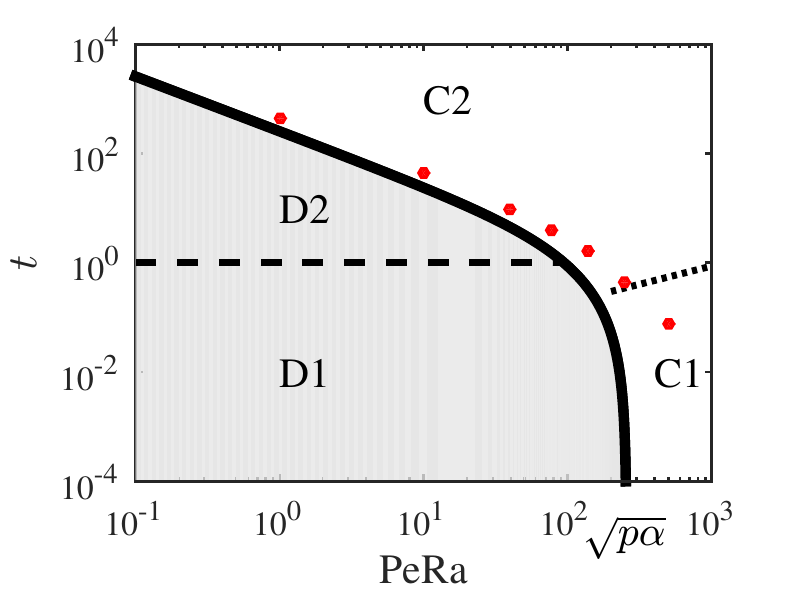}
	\caption{Diagram highlighting the different regimes of solute mass transport predicted by the 1D advection-dispersion model, Eqs.~(\ref{eq1Daverage}-\ref{limeq1Daverage}).  The thick line corresponds to  $t_{ \text{D2} \to \text{C2}}$  estimated from Eqs.~(\ref{eq:Deffapp}-\ref{eq:Phix0TEXT}). The shaded area  corresponds to the diffusive regimes D1 and D2.
	The dotted line corresponds to the transition between the dispersive  regimes C1 $\to$ C2, estimated from Eq.~(\ref{eq:tc1c2}).
	The red bullets represent the transition from a diffusive to a dispersive regime,  estimated with the 2D model, Eqs.~(\ref{equx}-\ref{eq:CLinfAdim}), 
	for $\mathrm{Pe} \simeq 0.252$ and $\mathrm{Ra}$ varying from $4$ to $2000$ (critical time estimated from Eq.~(\ref{eq:crit})).  
\label{fig:Diagramme}}
\end{center}
\end{figure}
 This  results in the diagram shown in Fig.~\ref{fig:Diagramme}, evidencing the transition from the diffusion-dominated regimes D1 and D2
towards the  dispersive  regime (C2). 

The 1D advection-dispersion model predicts that for $\text{Pe} \text{Ra} \geq \sqrt{p \alpha}$, buoyancy always dominates solute transport, regardless of the time (regimes C1 and C2).
In this range of parameters, 
we estimated the transition time $t_{\text{C1} \to \text{C2}}$  from the cross-over of $\delta$ vs. $t$ in both regimes C1 and C2, see Eq.~(\ref{eq:deltadiffbuo}) and Eq.~(\ref{eq:deltaC2}), leading to:
\begin{equation}
t_{\text{C1} \to \text{C2}} \simeq  0.6 \left(\frac{\text{Pe}\text{Ra}}{\sqrt{\alpha}}\right)^{2/3}. \label{eq:tc1c2} 
\end{equation} 
Figure~\ref{fig:Diagramme} summarizes all these results and can therefore serve as a guide to predict the expected transport regimes in a given experimental configuration using a single dimensionless parameter $\text{Pe}\text{Ra}$.

\subsection{Validity and limitations of the 1D advection-dispersion model \label{sec:validity}}

The 1D advection-dispersion model is based on several simplifying assumptions, whose validity must be carefully checked in order to validate the results in a given configuration. 

The assumptions required to validate Eq.~(\ref{eq:fondamentale}), on which the estimation of the dispersion term in Eq.~(\ref{eq:transpAverage}) is based, 
are the following ones (see Appendix~\ref{app:derivation}): 
\begin{eqnarray}
&&t \gg \text{Pe}^2,\label{eq:cond1} \\ 
&&\delta \gg \text{Pe}, \label{eq:cond2}\\ 
&&\text{Pe}^2 \varphi_1 \ll \varphi_0. \label{eq:cond3}
\end{eqnarray}
The last inequality Eq.~(\ref{eq:cond3}) demands that the transverse dispersion time $\delta^2/D_\text{eff}$ remains longer than the 
diffusion time across the channel height $\sim \text{Pe}^2$~\cite{Young1991}.
In addition, $\beta \text{Pe}^2 \ll 1$ is required to neglect the Taylor-Aris contribution in Eq.~(\ref{eq:Deffsansapp}). 
As  $\beta = 1/210$ ($1/\sqrt{\beta} \simeq 14.5$),
this last assumption is always verified for $\text{Pe} \lesssim 1$ (i.e. $\text{Pe} \ll 1$ or $\text{Pe} \sim 1$).
In the remainder of this analysis, we  will show that $\text{Pe} \lesssim 1$ results in the validity of conditions Eqs.~(\ref{eq:cond1}-\ref{eq:cond3}) 
from $t=0$ to $t\to \infty$ when $\text{PeRa} \ll \sqrt{\alpha}$ (asymptotic case 1 in the following analysis),  and from $t \gg \text{Pe}^2$ to $t\to \infty$ when $\text{PeRa} \gg \sqrt{\alpha}$  (asymptotic case 2).

\begin{figure}[ht]
\begin{center}
\includegraphics{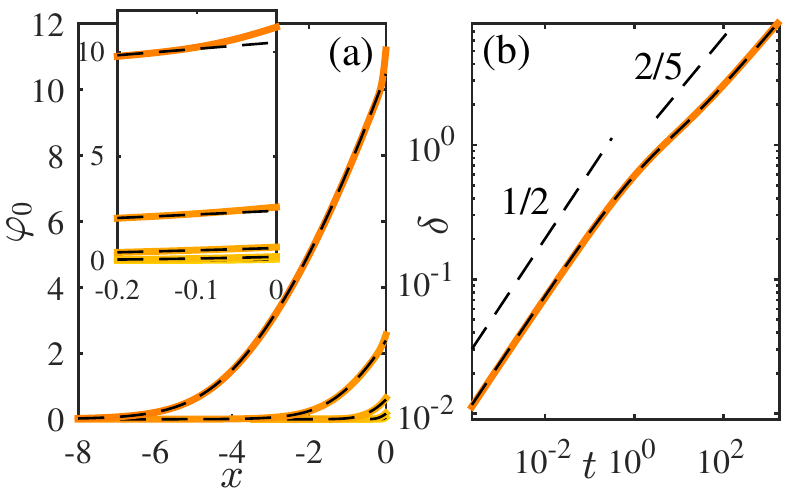}
\caption{(a) Average dimensionless concentration profiles $\varphi_0(x,t)$ for both the 2D model (same parameters as in Sec~\ref{experimentalcase} leading to
$\text{Pe} \simeq 0.344$ and $\text{Ra} \simeq 910$), 
and the solution of the advection-dispersion equation Eq.~(\ref{eq1Daverage}) for $\text{Pe}\text{Ra} \simeq 313$ (black dotted lines), 
at times $t\simeq 2.29 \times 10^{-3}$, $2.29 \times 10^{-2}$,  $0.229$, $2.29$ and $22.9$.
The inset shows a zoom close to the drying interface $x=0$.
(b) Extent of the diffusive layer $\delta$ defined by Eq.~(\ref{eq:defdelta}) for the 2D model and the 1D model (black dotted line) with the same parameters as in (a).}
\label{fig:FigureFD2}
\end{center}
\end{figure}

\paragraph{\textit{Case 1:}} $\text{PeRa} / \sqrt{\alpha} \ll 1$, corresponding to the succession of regimes D1, D2, C2,
with a transition from the diffusive regime D2 to the dispersive  regime C2 at time  $t_{ \text{D2} \to \text{C2}}$ given by Eq.~(\ref{eq:tD2C2}).
We show in appendix \ref{App:val} that conditions Eqs.~(\ref{eq:cond1}-\ref{eq:cond3}) are always satisfied 
for times $t \gg \, t_{ \text{D2} \to \text{C2}}$.
Therefore, the model is expected to describe correctly the   dispersive regime  C2.
Conditions Eqs.~(\ref{eq:cond1}-\ref{eq:cond3}) might be wrong in regimes D1 or D2 (i.e. for $t \ll t_{ \text{D2} \to \text{C2}}$).
But no significant loss of accuracy is expected, since buoyancy-driven dispersion is negligible in these diffusive regimes. 
Consequently, the 1D advection-dispersion model is valid at all times, from $t=0$ to $t \to \infty$. 
This is illustrated in Fig.~\ref{fig:FigureFD2}, where the 1D advection-dispersion model is compared 
with the output of the 2D model defined by Eqs.~(\ref{equx}-\ref{eq:CLinfAdim}), for $\text{Pe} \simeq 0.344$ and $\text{Ra} \simeq 910$, 
leading to $\text{PeRa}/\sqrt{\alpha} \simeq 0.52$.
The agreement is almost perfect for both $\varphi_0$ and $\delta$, from short to long times. 
An exception is the weak loss of accuracy of the 1D advection-dispersion model for the estimation of $\varphi_0$ close to the boundary at $x=0$ 
(see the inset in Fig.~\ref{fig:FigureFD2}a), 
because the lubrication theory does not fit the boundary condition Eq.~(\ref{eq:afreesurfaceAdim}). 
As expected, the discrepancy between both models is limited to a region whose extent is of the order of $\text{Pe}$
(the height of the slit  $h$ scaled by the reference horizontal length $D/\dot E$).

\paragraph{\textit{Case 2:}} $\text{PeRa} / \sqrt{\alpha} \gg 1$, corresponding to the succession of  the dispersive regimes C1 and C2.
We show in Appendix~\ref{App:val} that Eqs.~(\ref{eq:cond1}-\ref{eq:cond3}) are all verified as soon as  $t \gg \text{Pe}^2$.
The 1D advection-dispersion model must therefore be used carefully in this case, because this condition is not valid at the beginning,  
for short times since $\text{Pe} \lesssim 1$. This point is illustrated in Fig.~\ref{fig:Diagramme}, 
where the transition from the diffusive to the  dispersive  regimes has been determined with the 1D advection-dispersion model (thick continuous line) 
and with the 2D model defined by Eqs.~(\ref{equx}-\ref{eq:CLinfAdim}) (red bullets). In the latter case, we set $\text{Pe} \simeq 0.252$ and Ra has been varied from $4$ to $2000$.
The critical times estimated with the 2D model are close to the estimates given by the 1D advection-dispersion model, except for the two highest $\text{Pe}\text{Ra}$,
corresponding to the succession of the C1-C2 regimes in the 1D approach.
The 2D model shows that a diffusive regime always exists at small times,  
whereas the 1D model, which is not valid at short times $t \lesssim \text{Pe}^2$, erroneously predicts that the dispersive  regime begins from $t=0$.

\section{Conclusion and discussions \label{sec:conclusion}}

In the present work, we investigated the role of buoyancy-driven free convection on the solute mass transport 
in a model experimental configuration: solvent evaporation from a dilute mixture confined in  a horizontal slit. To quantify the impact of buoyancy, we derived a  1D advection-dispersion equation as traditionally done in the context of shear dispersion. This equation displays a dispersion coefficient $D_{\text{eff}}$ to account for buoyancy-driven flows, see Eq.~(\ref{eq:Deffdim}). Solute mass transport remains dominated by diffusion as long as:
\begin{eqnarray}
\left(\frac{g \beta_s h^4}{\nu D } \frac{\partial  \Phi_0}{\partial X } \right)^2\ll \alpha,  \label{ineq:alpha}
\end{eqnarray}
with real units, 
 where $\alpha = 362880$.    
We then performed a complete analysis of the expected regimes in the configuration shown in Fig.~\ref{fig:Evap1D}, and we showed in particular that 
the solutes transport is always dominated by buoyancy-driven dispersion  at long time scales leading to a continuously increasing diffusive layer as $\delta \propto T^{2/5}$.
A critical point of our model is the assumption  of constant mutual diffusion coefficient and viscosity during the solute concentration, and observing the different regimes predicted above maybe challenging  in a single experiment (see for instance the  case explored in Fig.~\ref{fig:FigureFD1} and the associated discussion). Eq.~(\ref{eq:Deffdim}) can be easily modified to include a slowing down of the evaporation rate and a non-constant mutual diffusion coefficient at high solute concentration, as done for instance in Ref.~\cite{Salmon2017} in a similar configuration for polymer and surfactant solutions, but without buoyancy. Nevertheless, the different regimes predicted above, and their associated scaling laws,  could be impacted by these effects. 

Note also that our work focused on the case of a slit, and we included in the above calculation the dispersion due to the Poiseuille flow, see the Taylor-Aris term in  Eq.~(\ref{eq:Deffsansapp}). Moreover, there are no coupling terms between buoyancy and Taylor-Aris dispersion owing to the symmetry of the geometry investigated, and we had therefore neglected the dispersion due to the Poiseuille flow. 
Interestingly, the case of a capillary tube with a finite width (as a square or a circular cross-section)
should deserve further attention. 
Indeed, a calculation similar to that performed above for such geometries can be carried out, but density gradients in the transverse dimension lead to transverse flows that couple to the longitudinal dispersion. To our knowledge, Chatwin and 
Erdogan~\cite{Erdogan1967} were the first to mention this subtle point when studying the dispersion of a solute flowing in a straight circular tube in the presence of buoyancy. They even showed that these  transverse flows could decrease the  buoyancy-induced longitudinal dispersion for a given range of parameters. For capillary tubes,   one cannot also 
exclude  (possibly non-negligible) coupling terms between Taylor-Aris and buoyancy-driven dispersions in the configuration shown in Fig.~\ref{fig:Evap1D}.  
We hope in a near future to investigate these subtle issues in detail.

Beyond the specific configuration investigated in the present work, different microfluidic tools have recently been developed to measure accurately the mutual diffusion coefficient of liquid mixtures owing to the very precise control of experimental conditions and mass transport at small scales~\cite{Vogus:15,Peters2017,Bouchaudy:18}.  In this context, Eq.~(\ref{eq:Deffdim}) makes it possible to rigorously estimate the impact of buoyancy on such measurements. More specifically, experiments exploiting drying to induce concentration gradients, similarly to the experiment shown in Fig.~\ref{fig:Evap1D}, were even recently reported to measure mutual diffusion coefficients of various complex fluids, namely copolymer solutions~\cite{Daubersies:12} and charged colloidal dispersions~\cite{Goehring2017,Loussert:16}. For the latter case, the rheological properties of the colloidal dispersions strongly evolve with the colloid (solute) concentration  up to reaching the formation of colloidal glasses at a concentration below the colloid close-packing. Nevertheless,  buoyancy-driven flows were clearly evidenced
in the liquid regime~\cite{Loussert:16}, and rough estimates using Eq.~(\ref{ineq:alpha}) clearly indicate that buoyancy plays an important role for mass transport, casting some doubts on the coefficient values reported in~\cite{Loussert:16} at low colloid concentrations.

Beyond these measurements of mutual diffusion coefficients, our work may  be also of interest to evaluate the impact of buoyancy-driven dispersion for any other microfluidic configuration generating concentration gradients. In particular, many recent  works focused on diffusio-phoresis, i.e. the transport of colloidal particles induced by solute concentration gradients~\cite{Abecassis:09,NeryAzevedo:17,Shin2016}. The role of buoyancy in such experiments was even investigated recently in detail by Gu {\it et al.} \cite{Gu2018}. Using scaling arguments, they identified conditions for which 
buoyancy negligibly impacts solute mass transport, leading to an inequality similar to Eq.~(\ref{ineq:alpha})
but with a significantly different numerical constant ($96^2 = 9216$ instead of $\alpha = 362880$ for a slit). Our theoretical development based on a standard pertubative approach~\cite{Young1991} may therefore help to refine the range of parameters for which buoyancy-driven dispersion  does not play any role in such experiments. Diffusio-phoretic effects are moreover 
 expected to play a crucial role in evaporating liquid mixtures of colloids of different sizes, possibly leading  to stratified materials~\cite{Sear2017}. 
We also hope that our simple model of uni-directional drying  may be relevant to evaluate the role of buoyancy in similar configurations, in particular when
gradients are orthogonal to gravity as in evaporation-induced propagating fronts in drying liquid films~\cite{Routh:13}.

 \begin{acknowledgments}
We thank Solvay and CNRS for fundings. 
\end{acknowledgments}

\appendix
\section{Derivation of the advection-dispersion equation \label{app:derivation}}

Using the dimensionless variables given by Eqs.~(\ref{eq:unitless}),  Eqs.~(\ref{eq:stokdim}-\ref{eq:convdim}) read:  
\begin{eqnarray}
&&\left(\text{Pe}^2\frac{\partial^2}{\partial x^2} + \frac{\partial^2}{\partial z^2}\right)u_x = \frac{\partial p}{\partial x}, \label{equx}\\
&&\text{Pe}^2\left(\text{Pe}^2\frac{\partial^2}{\partial x^2}  + \frac{\partial^2}{\partial z^2}  \right)u_z = \frac{\partial p}{\partial z}  + \text{Ra} \varphi, \label{equz}\\
&&\nabla.\mathbf{u}  = 0, \label{eq:contAdim} \\ 
&&\frac{\partial \varphi}{\partial t} + \mathbf{u}. \nabla \varphi = \left(\frac{\partial^2}{\partial x^2} + \frac{1}{\text{Pe}^2}\frac{\partial^2}{\partial z^2}\right) \varphi, \label{eq:convsansdim}
\end{eqnarray}
The dimensionless initial condition is $\varphi(x,z,t=0)=0$ and the dimensionless boundary conditions are:
\begin{eqnarray}
&&u_x(x=0,z,t) = 1, \label{eq:afreesurfaceAdim} \\
&&\left(\frac{\partial u_z}{\partial x}\right)(x=0,z,t)  =  0, \label{eq:bfreesurfaceAdim} \\
&&\left(1+ \varphi -  \frac{\partial \varphi}{\partial x}\right)(x=0,z,t) = 0, \label{eq:cfreesurfaceAdim} \\
&&\varphi(x \to -\infty ,z,t) = 0, \label{eq:CLinfAdim}
\end{eqnarray} 
along with the dimensionless counterpart of Eqs.~(\ref{eq:slip}-\ref{eq:adiab}) at the solid walls.

Averaging the transport equation~(\ref{eq:convsansdim}) over the height $h$ leads to,  with the help of Eq.~(\ref{eq:perturbation})
\begin{eqnarray}
&&\frac{\partial  \varphi_0}{\partial t} + \frac{\partial  \varphi_0}{\partial x}  + \text{Pe}^2 \frac{\partial  <u_x \varphi_1>}{\partial x}  = \frac{\partial^2  \varphi_0}{\partial x^2}  \label{eq:transpAverage}.
\end{eqnarray}
 Subtracting this last relation to Eq.~(\ref{eq:convsansdim}) results in:
\begin{multline}
\frac{\partial  \varphi_1}{\partial t} + \frac{(u_x-1)}{\text{Pe}^2} \frac{\partial \varphi_0}{\partial x} +  \mathbf{u}. \nabla \varphi_1 - \frac{\partial  <u_x \varphi_1>}{\partial x}  =\\ \frac{\partial^2  \varphi_1}{\partial x^2} + \frac{1}{\text{Pe}^2}\frac{\partial^2  \varphi_1}{\partial z^2}   \label{eq:transpAveragesub}.
\end{multline}
The continuity equation~Eq.~(\ref{eq:contAdim}) imposes the  scaling $u_z \sim u_x/\delta$, and Eq.~(\ref{eq:transpAveragesub})  therefore leads to:
\begin{eqnarray}
 \frac{\partial^2  \varphi_1}{\partial z^2}  \simeq (u_x-1) \frac{\partial \varphi_0}{\partial x}.  \label{eq:fondamentale}
\end{eqnarray}
assuming $t \gg \text{Pe}^2$, $\delta \gg \text{Pe}$, and $\text{Pe}^2 \varphi_1 \ll \varphi_0$~\cite{Young1991}.

Similarly, the leading-order terms in the Stokes equation~Eqs.~(\ref{equx}-\ref{equz}) lead to the horizontal component of the velocity field:
\begin{eqnarray}
&&u_x(x,z,t) =u_x^P(z) + u_x^B(x,z,t), \label{eq:vTABuo}\\
&&u_x^P(z) = 6 z (1-z)  , \notag \\
&&u_x^B(x,z,t) = -  \frac{\text{Ra}}{12}\frac{\partial  \varphi_0}{\partial  x} z (2z-1)(z-1), \notag  
\end{eqnarray}
checking both  the global mass balance Eq.~(\ref{eq:convglobaleadim}) and the no-slip boundary conditions on the solid walls.
The term $u_x^P$ is simply the Poiseuille flow induced by solvent evaporation, whereas the second term $u_x^B$, known as the Birikh profile \cite{BIRIKH1966}, corresponds to the flow induced by buoyancy, see Fig.~\ref{fig:Evap1D}(a).
Notice that the velocity field Eq.~(\ref{eq:vTABuo}) does not fit the boundary condition Eq.~(\ref{eq:afreesurfaceAdim}). 
This is a usual drawback of the lubrication theory which results in a loss of accuracy in the vicinity of the interface at $x=0$, see  
Sec.~\ref{sec:validity} for a discussion.

$\varphi_1$ can now be evaluated from Eq.~(\ref{eq:fondamentale}), assuming the no-flux boundary condition at $z=0$ and $z=1$, and imposing $<\varphi_1>=0$.
Using the linearity of Eq.~(\ref{eq:fondamentale}), we look separately for the solutions $\varphi_1^P$ and $\varphi_1^B$ due to the Poiseuille flow and buoyancy respectively, i.e.:
\begin{eqnarray}
&& \frac{\partial^2  \varphi_1^P}{\partial z^2}  = (u_x^{P} -1)\frac{\partial \varphi_0}{\partial x} \label{eq:phi12D_2}\\ 
&&\frac{\partial^2  \varphi_1^B}{\partial z^2}= u_x^{B}\frac{\partial \varphi_0}{\partial x}. \label{eq:PoissonBuo}
\end{eqnarray} 
After calculation, one finds:
 \begin{eqnarray}
&&\varphi_1(x,z,t) = \varphi_1^P(x,z,t) +\varphi_1^B(x,z,t), \label{eq:vp1}\\
&& \varphi_1^P(x,z,t) = \frac{\partial  \varphi_0}{\partial  x} \left( z^3 - \frac{z^4}{2}  - \frac{z^2}{2} + \frac{1}{60}\right), \notag \\
&&\varphi_1^B(x,z,t) = -\frac{\text{Ra}}{1440} \left(\frac{\partial  \varphi_0}{\partial  x}\right)^2 \left( 12z^5- 30 z^4 + 20z^3- 1\right).  \notag 
\end{eqnarray}
This relation combined with the velocity profile given by Eq.~(\ref{eq:vTABuo}) can now be used to calculate the dispersion term $<u_x \varphi_1>$ in Eq.~(\ref{eq:transpAverage}),
leading to Eqs.~(\ref{eq1Daverage}-\ref{eq:Deffsansapp}).

\section{Fedorchenko and Chernov analytical solution in the diffusive regimes \label{appsolutionanalytik}}
Fedorchenko and Chernov~\cite{Fedorchenko2003} derived the analytic solution of Eq.~(\ref{eq:transport1D}) with the initial and boundary conditions given by 
Eqs.~(\ref{eq:phicl2}-\ref{limeq1Daverage}):
\begin{multline} \label{eq:Phi0analytique}
\varphi_0(x,t) =  \sqrt{\frac{t}{\pi}} \exp\left(-\frac{(t-x)^2}{4t}\right) \\ 
+ \frac{1}{2}\left\{\exp(x) (1+x+t)\text{erfc}\left(\frac{-(t+x)}{2\sqrt{t}}\right)  -\text{erfc}\left(\frac{t-x}{2\sqrt{t}}\right) \right\}.
\end{multline}
The concentration gradient at the interface, used in Sec.~\ref{TransTime}, simply follows from the spatial derivation of Eq.~(\ref{eq:Phi0analytique}) at $x=0$:
\begin{multline} \label{eq:Phix0TEXT}
\left( \frac{\partial \varphi_0}{\partial x} \right)_{x=0,~t} = \sqrt{\frac{t}{\pi}}  \exp \left (-\frac{t}{4} \right ) \\
+   \left ( 1+\frac{t}{2} \right )   \mathrm{erfc} \left ( - \frac{\sqrt t }{2} \right). 
\end{multline}
From Eq.~(\ref{eq:Phi0analytique}), one can also calculate the extent of the diffusive layer using Eq.~(\ref{eq:defdelta}), see Eq.~(\ref{eq:analdelta}).
 
\section{Approximate solutions using the integral method in the dispersive  regime C2 \label{appMethoInt}}
At long time scales, the temporal and diffusive terms in Eq.~(\ref{eq:phigov}) are negligible in the regime C2, and concentration profiles obey the following partial differential equation in the growing diffusive layer:
\begin{eqnarray}
 \frac{\partial \varphi_0}{\partial x}  \simeq \frac{(\text{Pe}\text{Ra})^2}{\alpha} \frac{\partial}{\partial x}\left( \frac{\partial \varphi_0}{\partial x} \right)^3.  \label{eq:B1}
\end{eqnarray}  
We define $\psi = \left(\frac{\partial \varphi_0}{\partial x} \right)^2$, and Eq.~(\ref{eq:B1}) becomes:
\begin{eqnarray}
\left(\frac{3}{2}\frac{(\text{Ra}\text{Pe})^2}{\alpha} \frac{\partial \psi}{\partial x}  - 1 \right) \psi^{1/2} \simeq 0.
\end{eqnarray} 
As the concentration gradient steadily increases, $\psi^{1/2} \neq 0$, and one has thus:
\begin{eqnarray}
 \frac{\partial \psi}{\partial x} \simeq \frac{2}{3}\frac{\alpha}{(\text{Ra}\text{Pe})^2},
\end{eqnarray} 
leading after integration to 
\begin{eqnarray}
\varphi_0(x,t) \simeq \sqrt{\frac{8{\alpha}}{27}} \frac{1}{\text{Pe} \text{Ra}}(x + G(t))^{3/2} + F(t),
\end{eqnarray} 
where $F$ and $G$ are two functions to be defined. 
We postulate  following the integral method~\cite{Ozisik} that the  relation:
\begin{eqnarray} \label{eq:appIPhi}
&&\varphi_0(x,t) = 0~\text{for~}x<-G(t),\\ 
&&\varphi_0(x,t) = \sqrt{\frac{8{\alpha}}{27}} \frac{1}{\text{Pe} \text{Ra}}(x + G(t))^{3/2} + F(t)~\text{for~}x<-G(t),  \notag
\end{eqnarray} 
 is a good approximation of the solution providing that it verifies both the boundary condition Eq.~(\ref{limeq1Daverage}) and the global solute conservation Eq.~(\ref{eq:convglobalsoluteeadim}).
 which leads to $F \simeq -1$, and:
\begin{eqnarray}
G(t) = \left(\frac{5}{2} \sqrt{\frac{27}{8 \alpha}} \text{Pe} \text{Ra} \, t \right)^{2/5}.  
\end{eqnarray} 
As shown in Fig.~\ref{fig:Evap1Deff_pdepe}, this relation approximates well the concentration profiles in the  dispersive  regime. From this approximation, we can finally calculate $\delta(t)$ using Eq.~(\ref{eq:defdelta}) leading to Eq.~(\ref{eq:deltadiffbuo}).

\section{Validity of the 1D advection dispersion model \label{App:val}}

We consider the two cases defined in section \ref{sec:validity}. 
\paragraph{\textit{Case 1:}} $\text{PeRa} / \sqrt{\alpha} \ll 1$. 
We aim at demonstrating that if  $\text{Pe} \lesssim 1$, then conditions Eqs.~(\ref{eq:cond1}-\ref{eq:cond3}) hold for time 
$t \gg t_{ \text{D2} \to \text{C2}}$, 
where $t_{ \text{D2} \to \text{C2}}$ is given by Eq.~(\ref{eq:tD2C2}).
The condition Eq.~(\ref{eq:cond1}) is satisfied because $t \gg t_{ \text{D2} \to \text{C2}} \gg 1$ and $\text{Pe} \lesssim 1$.
Similarly, the condition Eq.~(\ref{eq:cond2}) is true because $t \gg t_{ \text{D2} \to \text{C2}}$ implies $\delta \gg 1$ (see scalings~(\ref{eq:C2}-\ref{eq:C2bis})).
Using Eq.~(\ref{eq:vp1}) and assuming $1440 \sim \sqrt{\alpha}$, the condition Eq.~(\ref{eq:cond3}) reads 
\begin{equation}
\text{Pe} \, \frac{\text{PeRa}}{\sqrt{\alpha}} \, \frac{\varphi_0}{\delta^2} \ll 1 \label{eq:cond3bis} \, .
\end{equation}
Using scalings~(\ref{eq:C2}-\ref{eq:C2bis}), 
Eq.~(\ref{eq:cond3bis}) reduces to 
\begin{equation}
\text{Pe}  \left ( \frac{\text{PeRa}}{\sqrt{\alpha}} \, t \right ) ^{-1/5} \ll 1,  \label{eq:cond3ter}
\end{equation}
this condition being obviously true for $t \gg t_{ \text{D2} \to \text{C2}}$.

\paragraph{\textit{Case 2:}} $\text{PeRa} / \sqrt{\alpha} \gg 1$. 
We aim at demonstrating that if condition Eq.~(\ref{eq:cond1}) is true, then conditions Eqs.~(\ref{eq:cond2}-\ref{eq:cond3}) are also true.
$\delta \gg \text{Pe} \, \left ( \text{PeRa}/\sqrt{\alpha} \right )^{1/3} \gg \text{Pe}$ in the C1 regime (from scaling (\ref{eq:C1}) with $t \gg \text{Pe}^2$), 
and $\delta  \gg 1$ in the C2 regime (from scaling (\ref{eq:C2}-\ref{eq:C2bis})), which proves the validity of condition (\ref{eq:cond2}).
In the C1 regime, condition (\ref{eq:cond3}) turns to Eq.~(\ref{eq:cond3bis}) and then to $\text{Pe} \, t^{-1/2} \ll 1$ (using scaling~(\ref{eq:C1})), 
which is true for $t \gg \text{Pe}^2$.
In the C2 regime, condition~(\ref{eq:cond3}) still leads to Eqs.~(\ref{eq:cond3bis}) and (\ref{eq:cond3ter}), the latter being true for $t \gg \text{Pe}^2$. \\

\end{document}